\documentclass[12pt,a4paper,makeidx]{article}
\usepackage[latin1]{inputenc}
\usepackage{amsmath}
\usepackage{amsfonts}
\usepackage{amssymb}
\usepackage{epsfig}
\usepackage{theorem}
\usepackage{oldgerm}
\usepackage{tabularx}
\usepackage{placeins}
\usepackage{amscd}
\usepackage{bm}
\usepackage{epic}
\usepackage{curves}
\usepackage{graphicx}
\usepackage{makeidx}
\usepackage{fancyhdr}
\usepackage{graphpap}
\usepackage{rotating}
\usepackage{bbm}
\usepackage{pstricks}
\usepackage[absolute]{textpos}
\usepackage{psfrag}
\usepackage{times}
\usepackage{dcolumn}
\usepackage{shadow}
\usepackage{multicol}
\usepackage[latin1]{inputenc}
\usepackage{oldgerm}
\usepackage[left=3.5cm,top=4cm,right=3.5cm,bottom=4cm]{geometry}
\fancypagestyle{plain}{%
\fancyhead{} 
}

\makeatletter
\def\thebibliography#1{\centering {\section*{References\@mkboth
  {REFERENCES}{REFERENCES}}}\list
  {[\arabic{enumi}]}{\settowidth\labelwidth{[#1]}\leftmargin\labelwidth
  \advance\leftmargin\labelsep
	\usecounter{enumi}}
	\def\newblock{\hskip .11em plus .33em minus .07em}
	\sloppy\clubpenalty4000\widowpenalty4000
	\sfcode`\.=1000\relax}
\makeatother

\begin{document}
\vspace*{3.5cm}
\begin{center}
\textbf{IRREDUCIBLE REPRESENTATIONS OF THE $\bold{CPT}$ GROUPS IN QED} \\
\vspace{1cm}
B. Carballo Pérez$^{1}$, M. Socolovsky$^{2}$ \\
\small{ Instituto de Ciencias Nucleares, Universidad Nacional Aut\'onoma de M\'exico,}\\
\small{ Circuito exterior, Ciudad Universitaria, 04510, M\'exico D.F., México}\\
$^{1}$e-mail: brendacp@nucleares.unam.mx\\
$^{2}$e-mail: socolovs@nucleares.unam.mx
\end{center}
\vspace{0.5cm}
\textbf{Abstract:} We construct the inequivalent irreducible representations (IIR's) of the $CPT$ groups of the Dirac field operator $\hat{\psi}$ and the electromagnetic quantum potential $\hat{A_{\mu}}$. The results are valid both for free and interacting (QED) fields. Also, and for the sake of completeness, we construct the IIR's of the $CPT$ group of the Dirac equation.\\
\vspace{0.5cm}
\vspace{0.5cm}

\newpage

{\centering \section{Introduction}}

\hspace{18pt} The $CPT$ group $G_{\hat{\Theta}}(\hat{\psi})$ of the Dirac quantum field $\hat{\psi}$ is isomorphic to the direct product of the quaternion group $Q$ and the cyclic group of two elements $\mathbb{Z}_{2}$ (\cite{CPT group} and \cite{B1}):

\begin{equation}
G_{\hat{\Theta}}(\hat{\psi})\cong Q \times \mathbb{Z}_{2}.
\end{equation}

$\hat{\Theta}$ is the product $\hat{C}*\hat{P}*\hat{T}$ of the three operators: $\hat{C}$ (charge conjugation), $\hat{P}$ (space inversion), $\hat{T}$ (time reversal); the operation $\hat{A}*\hat{B}$, where $\hat{A}$ and $\hat{B}$ are any of the operators $\hat{C}$, $\hat{P}$, $\hat{T}$, is given by $(\hat{A}*\hat{B})\cdot\hat{\psi}=(\hat{A}\hat{B})^{\dagger}\hat{\psi}(\hat{A}\hat{B})$. $G_{\hat{\Theta}}(\hat{\psi})$ is one of the nine non abelian groups of a total of fourteen groups with sixteen elements \cite{Asche}; only three of them have three generators. 

The isomorphism $G_{\hat{\Theta}}(\hat{\psi}) \to Q \times \mathbb{Z}_2$ is given by the relations (\ref{Isomorfismo1}):

\begin{eqnarray}
1\mapsto (1,1) \qquad  -1\mapsto (-1,1)\nonumber\\ 
\hat{C}\mapsto (1,-1)  \qquad   -\hat{C}\mapsto (-1,-1)\nonumber\\
\hat{P}\mapsto (\iota,1)  \qquad  -\hat{P}\mapsto (-\iota,1)\nonumber\\
\hat{T}\mapsto (\gamma,1)  \qquad  -\hat{T}\mapsto (-\gamma,1)\nonumber\\
\hat{C}*\hat{P}\mapsto (\iota,-1)  \qquad  -\hat{C}*\hat{P}\mapsto (-\iota,-1)\nonumber\\
\hat{C}*\hat{T}\mapsto (\gamma,-1)  \qquad  -\hat{C}*\hat{T}\mapsto (-\gamma,-1)\nonumber\\
\hat{P}*\hat{T}\mapsto (\kappa,1)   \qquad   -\hat{P}*\hat{T}\mapsto (-\kappa,1)\nonumber\\
\hat{\Theta} \mapsto (\kappa,-1) \qquad  -\hat{\Theta} \mapsto (-\kappa,-1);
\label{Isomorfismo1}
\end{eqnarray}
where $\iota$, $\gamma$, $\kappa$ are the three imaginary units defining the quaternion numbers.

On the other hand, the action of the $\hat{C}$, $\hat{P}$ and $\hat{T}$ operators on the Maxwell electromagnetic 4-potential $\hat{A_{\mu}}$ is given by (\ref {actionA}) (\cite{Greiner} and \cite{Azcarraga}), with the  Minkowski space-time metric $diag (1,-1,-1,-1)$:

\begin{eqnarray}
\hat{P}\hat{A}^{\mu}(\bold{x},t)\hat{P}^{-1}=\hat{A}_{\mu}(-\bold{x},t),\nonumber\\
\hat{C}\hat{A}^{\mu}(\bold{x},t)\hat{C}^{-1}=-\hat{A}^{\mu}(\bold{x},t),\nonumber\\
\hat{T}\hat{A}^{\mu}(\bold{x},t)\hat{T}^{-1}=\hat{A}_{\mu}(\bold{x},-t).
\label{actionA}
\end{eqnarray}

This leads to the $CPT$ group of the electromagnetic field operator, $G_{\hat{\Theta}}(\hat{A_{\mu}})$, which is an abelian group of eight elements with three generators, isomorphic to $\mathbb{Z}_2\times\mathbb{Z}_2\times\mathbb{Z}_2=\mathbb{Z}_2^{3}$. Geometrically, $\mathbb{Z}_2^{3}$ is isomorphic to the group $D_{2h}$, the symmetry group of the parallelepiped.

The isomorphism
\begin{equation}
G_{\hat{\Theta}}(\hat{A_{\mu}})\rightarrow \mathbb{Z}_2\times\mathbb{Z}_2\times\mathbb{Z}_2
\end{equation}
is given by
\begin{eqnarray}
1\mapsto (e_{1},e_{2},e_{3}) \nonumber\\ 
\hat{C}\mapsto (a_{1},e_{2},e_{3})\nonumber\\
\hat{P}\mapsto (e_{1},a_{2},e_{3})\nonumber\\
\hat{T}\mapsto (e_{1},a_{2},a_{3})\nonumber\\
\hat{P}*\hat{T}\mapsto (e_{1},a_{2},a_{3})\nonumber\\
\hat{C}*\hat{P}\mapsto (a_{1},a_{2},e_{3})\nonumber\\
\hat{C}*\hat{T}\mapsto (a_{1},e_{2},a_{3})\nonumber\\
\hat{\Theta}\mapsto (a_{1},a_{2},a_{3})
\label{Isomorfismo2};
\end{eqnarray}
where $e_{i}$ and $a_{i}$ (for $i=1,2,3$) represent, respectively, the identity and the ``minus 1" of each of the $\mathbb{Z}_2$ groups.

As the $CPT$ transformation properties of the interacting $\hat{\psi}-\hat{A_{\mu}}$ fields are the same as for the free fields (\cite{Azcarraga} and \cite{BandD}), it is then clear that the complete $CPT$ group for QED, $G_{\hat{\Theta}}(QED)$, is the direct product of the two above mentioned groups, $G_{\hat{\Theta}}(\hat{\psi})$ and $G_{\hat{\Theta}}(\hat{A_{\mu}})$, i.e.,

\begin{equation}
G_{\hat{\Theta}}(QED)=G_{\hat{\Theta}}(\hat{\psi})\times G_{\hat{\Theta}}(\hat{A_{\mu}}),
\end{equation}
which turns out to be a group of order $\vert G_{\hat{\Theta}}(QED) \vert=16 \times 8=128$.

Thus,
\begin{equation}
G_{\hat{\Theta}}(QED) \cong (Q \times \mathbb{Z}_{2})\times \mathbb{Z}_{2}^{3}.
\end{equation}

Since, as will be shown below, $G_{\hat{\Theta}}(\hat{\psi})$ has ten inequivalent irreducible representations (IIR's), while $G_{\hat{\Theta}}(\hat{A_{\mu}})$ has only eight IIR's, then, the total number of IIR's of $G_{\hat{\Theta}}(QED)$ is eighty: sixty four of them 1-dimensional and sixteen 2-dimensional. 

It is interesting at this point to make a comment about the geometrical content of $G_{\hat{\Theta}}(QED)$, which is a set of points in spheres. In fact, each factor $\mathbb{Z}_{2}$ is a 0-sphere $S^{0}$, while $Q$ can be thought as a subset of eight points in the 3-sphere $S^{3}$. Thus, topologically

\begin{equation}
G_{\hat{\Theta}}(QED)\subset SU(2)\times (U(1))^{4}.
\end{equation}

In section 2 we construct by tensor products the ten IIR's and character tables of $G_{\hat{\Theta}}(\hat{\psi})$. In section 3 we construct the eight IIR's of $G_{\hat{\Theta}}(\hat{A_{\mu}})$; in this case, since all representations are 1-dimensional, the characters coincide with the representations. In section 4 we construct the IIR's of the $CPT$ group corresponding to the Dirac equation for the wave function $\psi$, $G_{\Theta}^{(2)}(\psi)$ in \cite{CPT group}, and compare the result with the case for the Dirac field operator $\hat{\psi}$. Finally, in section 5, we discuss the relation between the ¨fermionic" and ¨bosonic" $CPT$ groups and comment about the relation of these groups with some approaches to quantum paradoxes.

{\centering \section{IIR's of $\mathbf{G_{\hat{\bm \Theta}}(\hat{\bm \psi})}$}}

\hspace{18pt} The irreducible representations (irrep's) of the direct product of two groups, $G\times H$, are the tensor products of the irrep's of each of the factors, namely the set $\lbrace r_{G}\otimes r_{H}\rbrace$ \footnote{$\otimes=\otimes_{\mathbb{C}}$ is the tensor product over the complex numbers}  for all $r_{G}'s$, irrep's of $G$ and all $r_{H}'s$, irrep's of $H$ \cite{Sternberg}. If $G$ is the quaternion group, which, as is well known, has five IIR's $\varphi_{\mu}$ (with $\varphi_{1}, ..., \varphi_{4}$ 1-dimensional and $\varphi_{5}$ 2-dimensional)\cite{Hamermesh} and characters table given by table \ref{ChQ}; and $H$ is $\mathbb{Z}_{2}$ (with IIR's $\psi_{1}$ and $\psi_{2}$), then $G_{\hat{\Theta}}(\hat{\psi})$ has ten IIR's:

\begin{eqnarray}
\phi_{\alpha}&=&\varphi_{\alpha}\otimes\psi_{1},\quad\phi_{\alpha+4}=\varphi_{\alpha}\otimes\psi_{2},\quad \alpha=1,2,3,4,\\
\phi_{9}&=&\varphi_{5}\otimes\psi_{1},\quad\phi_{10}=\varphi_{5}\otimes\psi_{2}.
\end{eqnarray}
$\phi_{1},...,\phi_{8}$ are 1-dimensional, while $\phi_{9}$ and $\phi_{10}$ are 2-dimensional.

\begin{table}[h]
\begin{center}
\begin{tabular}{ccccccc}
\hline
\hline
Ch Q & \vline \vline & $[1]$ & $[-1]$ & $2[\iota]$ & $2[\gamma]$ & $2[\kappa]$ \\
\hline
\hline
$\chi_{1}$ & \vline \vline & $1$ & $1$ & $1$ & $1$ & $1$  \\
$\chi_{2}$ & \vline \vline & $1$ & $1$ & $1$ & $-1$ & $-1$  \\
$\chi_{3}$ & \vline \vline & $1$ & $1$ & $-1$ & $1$ & $-1$\\
$\chi_{4}$ & \vline \vline & $1$ & $1$ & $-1$ & $-1$ & $1$\\
$\chi_{5}$ & \vline \vline & $2$ & $-2$ & $0$ & $0$ & $0$\\
\hline
\hline
\end{tabular}
\end{center}
\caption[]
{\small Characters table of Q.}
\label{ChQ}
\end{table}

The corresponding characters are the functions
\begin{eqnarray}
\kappa_{\alpha}&=&\chi_{\alpha}\varphi_{1}:\;Q\times\mathbb{Z}_{2}\rightarrow\mathbb{C},\quad\kappa_{\alpha}(q,h)=\chi_{\alpha}(q)\varphi_{1}(h),\\
\kappa_{\alpha+4}&=&\chi_{\alpha}\varphi_{2}:\;Q\times\mathbb{Z}_{2}\rightarrow\mathbb{C},\quad\kappa_{\alpha+4}(q,h)=\chi_{\alpha}(q)\varphi_{2}(h),\\
\kappa_{9}&=&\chi_{5}\varphi_{1}:\;Q\times\mathbb{Z}_{2}\rightarrow\mathbb{C},\quad\kappa_{9}(q,h)=\chi_{5}(q)\varphi_{1}(h),\\
\kappa_{10}&=&\chi_{5}\varphi_{2}:\;Q\times\mathbb{Z}_{2}\rightarrow\mathbb{C},\quad\kappa_{10}(q,h)=\chi_{5}(q)\varphi_{2}(h).
\end{eqnarray}

These representations and characters (for conjugate classes) are explicitly given by the tables \ref{TID}
and \ref{TChD}, respectively, where $\lambda_{i}=tr\phi_{i}$, and the conjugate classes are given by the relations (\ref{cg1}):

\begin{table}[h]
\begin{center}
\begin{sideways}
\resizebox{1.5\textwidth}{!}{
\begin{tabular}{cccccccccccccccccccc}
\hline
\hline
IIR's & \vline \vline & $\hat{I}$ & $\hat{C}$ & $-\hat{I}$ & $-\hat{C}$ & $\hat{P}$ & $\hat{C}*\hat{P}$ & $-\hat{P}$ & $-\hat{C}*\hat{P}$ & $\hat{T}$ & $\hat{C}*\hat{T}$ & $-\hat{T}$ & $-\hat{C}*\hat{T}$ & $\hat{P}*\hat{T}$ & $\hat{\Theta}$ & $-\hat{P}*\hat{T}$ & $-\hat{\Theta}$\\
$G_{\hat{\Theta}}(\hat{\psi})$ & \vline \vline & $(1,e)$ & $(1,a)$ & $(-1,e)$ & $(-1,a)$ & $(\iota,e)$ & $(\iota,a)$ & $(-\iota,e)$ & $(-\iota,a)$ & $(\gamma,e)$ & $(\gamma,a)$ & $(-\gamma,e)$ & $(-\gamma,a)$ & $(\kappa,e)$ & $(\kappa,a)$ & $(-\kappa,e)$ & $(-\kappa,a)$\\
\hline
\hline
$\phi_{1}$ & \vline \vline & $1$ & $1$ & $1$ & $1$ & $1$ & $1$ & $1$ & $1$ & $1$ & $1$ & $1$ & $1$ & $1$ & $1$ & $1$ & $1$\\
$\phi_{2}$ & \vline \vline & $1$ & $-1$ & $1$ & $-1$ & $1$ & $-1$ & $1$ & $-1$ & $1$ & $-1$ & $1$ & $-1$ & $1$ & $-1$ & $1$ & $-1$ \\
$\phi_{3}$ & \vline \vline & $1$ & $1$ & $1$ & $1$ & $1$ & $1$ & $1$ & $1$ & $-1$ & $-1$ & $-1$ & $-1$ & $-1$ & $-1$ & $-1$ & $-1$\\
$\phi_{4}$ & \vline \vline & $1$ & $-1$ & $1$ & $-1$ & $1$ & $-1$ & $1$ & $-1$ & $-1$ & $1$ & $-1$ & $1$ & $-1$ & $1$ & $-1$ & $1$\\
$\phi_{5}$ & \vline \vline & $1$ & $1$ & $1$ & $1$ & $-1$ & $-1$ & $-1$ & $-1$ & $1$ & $1$ & $1$ & $1$ & $-1$ & $-1$ & $-1$ & $-1$\\
$\phi_{6}$ & \vline \vline & $1$ & $-1$ & $1$ & $-1$ & $-1$ & $1$ & $-1$ & $1$ & $1$ & $-1$ & $1$ & $-1$ & $-1$ & $1$ & $-1$ & $1$\\
$\phi_{7}$ & \vline \vline & $1$ & $1$ & $1$ & $1$ & $-1$ & $-1$ & $-1$ & $-1$ & $-1$ & $-1$ & $-1$ & $-1$ & $1$ & $1$ & $1$ & $1$\\
$\phi_{8}$ & \vline \vline & $1$ & $-1$ & $1$ & $-1$ & $-1$ & $1$ & $-1$ & $1$ & $-1$ & $1$ & $-1$ & $1$ & $1$ & $-1$ & $1$ & $-1$\\
$\phi_{9}$ & \vline \vline & $\begin{pmatrix} 1 & 0 \cr 0 & 1 \end{pmatrix}$ & $\begin{pmatrix} 1 & 0 \cr 0 & 1 \end{pmatrix}$ & $\begin{pmatrix} -1 & 0 \cr 0 & -1 \end{pmatrix}$ & $\begin{pmatrix} -1 & 0 \cr 0 & -1 \end{pmatrix}$ & $\begin{pmatrix} i & 0 \cr 0 & -i \end{pmatrix}$ & $\begin{pmatrix} i & 0 \cr 0 & -i \end{pmatrix}$ & $\begin{pmatrix} -i & 0 \cr 0 & i \end{pmatrix}$ & $\begin{pmatrix} -i & 0 \cr 0 & i \end{pmatrix}$ & $\begin{pmatrix} 0 & 1 \cr -1 & 0 \end{pmatrix}$ & $\begin{pmatrix} 0 & 1 \cr -1 & 0 \end{pmatrix}$ & $\begin{pmatrix} 0 & -1 \cr 1 & 0 \end{pmatrix}$ & $\begin{pmatrix} 0 & -1 \cr 1 & 0 \end{pmatrix}$ & $\begin{pmatrix} 0 & i \cr i & 0 \end{pmatrix}$ & $\begin{pmatrix} 0 & i \cr i & 0 \end{pmatrix}$ & $\begin{pmatrix} 0 & -i \cr -i & 0 \end{pmatrix}$ & $\begin{pmatrix} 0 & -i \cr -i & 0 \end{pmatrix}$\\
$\phi_{10}$ & \vline \vline & $\begin{pmatrix} 1 & 0 \cr 0 & 1 \end{pmatrix}$ & $\begin{pmatrix} -1 & 0 \cr 0 & -1 \end{pmatrix}$ & $\begin{pmatrix} -1 & 0 \cr 0 & -1 \end{pmatrix}$ & $\begin{pmatrix} 1 & 0 \cr 0 & 1 \end{pmatrix}$ & $\begin{pmatrix} i & 0 \cr 0 & -i \end{pmatrix}$ & $\begin{pmatrix} -i & 0 \cr 0 & i \end{pmatrix}$ & $\begin{pmatrix} -i & 0 \cr 0 & i \end{pmatrix}$ & $\begin{pmatrix} i & 0 \cr 0 & -i \end{pmatrix}$ & $\begin{pmatrix} 0 & 1 \cr -1 & 0 \end{pmatrix}$ & $\begin{pmatrix} 0 & -1 \cr 1 & 0 \end{pmatrix}$ & $\begin{pmatrix} 0 & -1 \cr 1 & 0 \end{pmatrix}$ & $\begin{pmatrix} 0 & 1 \cr -1 & 0 \end{pmatrix}$ & $\begin{pmatrix} 0 & i \cr i & 0 \end{pmatrix}$ & $\begin{pmatrix} 0 & -i \cr -i & 0 \end{pmatrix}$ & $\begin{pmatrix} 0 & -i \cr -i & 0 \end{pmatrix}$ & $\begin{pmatrix} 0 & i \cr i & 0 \end{pmatrix}$\\
\hline
\hline
\end{tabular}}
\end{sideways}
\end{center}
\caption[]{\small{IIR's table of $G_{\hat{\Theta}}(\hat{\psi})$.}}
\label{TID}
\end{table}

\begin{table}[h]
\begin{center}
\resizebox{1.0\textwidth}{!}{
\begin{tabular}{cccccccccccccccccccc}
\hline
\hline
Ch $G_{\hat{\Theta}}(\hat{\psi})$ & \vline \vline & $[(1,e)]$ & $[(1,a)]$ & $[(-1,e)]$ & $[(-1,a)]$ & $2[(\iota,e)]$ & $2[(\iota,a)]$ & $2[(\gamma,e)]$ & $2[(\gamma,a)]$ &  2[$(\kappa,e)]$ & $2[(\kappa,a)]$\\
\hline
\hline
$\lambda_{1}$ & \vline \vline & $1$ & $1$ & $1$ & $1$ & $1$ & $1$ & $1$ & $1$ & $1$ & $1$\\
$\lambda_{2}$ & \vline \vline & $1$ & $-1$ & $1$ & $-1$ & $1$ & $-1$ & $1$ & $-1$ & $1$ & $-1$\\
$\lambda_{3}$ & \vline \vline & $1$ & $1$ & $1$ & $1$ & $1$ & $1$ & $-1$ & $-1$ & $-1$ & $-1$\\
$\lambda_{4}$ & \vline \vline & $1$ & $-1$ & $1$ & $-1$ & $1$ & $-1$ & $-1$ & $1$ & $-1$ & $1$\\
$\lambda_{5}$ & \vline \vline & $1$ & $1$ & $1$ & $1$ & $-1$ & $-1$ & $1$ & $1$ & $-1$ & $-1$\\
$\lambda_{6}$ & \vline \vline & $1$ & $-1$ & $1$ & $-1$ & $-1$ & $1$ & $1$ & $-1$ & $-1$ & $1$\\
$\lambda_{7}$ & \vline \vline & $1$ & $1$ & $1$ & $1$ & $-1$ & $-1$ & $-1$ & $-1$ & $1$ & $1$\\
$\lambda_{8}$ & \vline \vline & $1$ & $-1$ & $1$ & $-1$ & $-1$ & $1$ & $-1$ & $1$ & $1$ & $-1$\\
$\lambda_{9}$ & \vline \vline & $2$ & $2$ & $-2$ & $-2$ & $0$ & $0$ & $0$ & $0$ & $0$ & $0$\\
$\lambda_{10}$ & \vline \vline & $2$ & $-2$ & $-2$ & $2$ & $0$ & $0$ & $0$ & $0$ & $0$ & $0$\\
\hline
\hline
\end{tabular}}
\end{center}
\caption[]{\small{Characters table of $G_{\hat{\Theta}}(\hat{\psi})$.}}
\label{TChD}
\end{table}

\begin{eqnarray}
\text{$[$}\hat{I}]&=&[(1,e)]=\lbrace(1,e)\rbrace,\nonumber\\
\text{$[$}\hat{C}]&=&[(1,a)]=\lbrace(1,a)\rbrace,\nonumber\\
\text{$[$}-\hat{I}]&=&[(-1,e)]=\lbrace(-1,e)\rbrace,\nonumber\\
\text{$[$}-\hat{C}]&=&[(-1,a)]=\lbrace(-1,a)\rbrace,\nonumber\\
\text{$[$}\hat{P}]&=&[(\iota,e)]=\lbrace(\iota,e),(-\iota,e)\rbrace,\nonumber\\
\text{$[$}\hat{C}*\hat{P}]&=&[(\iota,a)]=\lbrace(\iota,a),(-\iota,a)\rbrace,\nonumber\\
\text{$[$}\hat{T}]&=&[(\gamma,e)]=\lbrace(\gamma,e),(-\gamma,e)\rbrace,\nonumber\\
\text{$[$}\hat{C}*\hat{T}]&=&[(\gamma,a)]=\lbrace(\gamma,a),(-\gamma,a)\rbrace,\nonumber\\
\text{$[$}\hat{P}*\hat{T}]&=&[(\kappa,e)]=\lbrace(\kappa,e),(-\kappa,e)\rbrace,\nonumber\\
\text{$[$}\hat{\Theta}]&=&[(\kappa,a)]=\lbrace(\kappa,a),(-\kappa,a)\rbrace;
\label{cg1}
\end{eqnarray}
where $e$ and $a$ are the elements of $\mathbb{Z}_{2}$.

The character table of $G_{\hat{\Theta}}(\hat{\psi})$ is the same as the character table of $D_{4h}$, the group of symmetries of a prism with square base, though $Q\times \mathbb{Z}_{2}\ncong D_{4h}$.

{\centering \section{IIR's of $\mathbf{G_{\hat{\bm \Theta}}(\hat{\bm A_{\bm \mu}})}$}}

\hspace{18pt} Since $G_{\hat{\Theta}}(\hat{A_{\mu}})\cong\mathbb{Z}_{2}^{3}$ is an abelian group, all its irrep's are 1-dimensional. Since $\vert G_{\hat{\Theta}}(\hat{A_{\mu}}) \vert=8$, then $G_{\hat{\Theta}}(\hat{A_{\mu}})$ has eight IIR's, which can be identified with the corresponding characters. Using the isomorphism between $G_{\hat{\Theta}}(\hat{A_{\mu}})$ and $\mathbb{Z}_{2}^{3}$ (eq. (\ref{Isomorfismo2})), and the character table of $\mathbb{Z}_{2}$, we obtain the table of representations ($\phi_{ijk}$) or characters ($\chi_{ijk}$) for $G_{\hat{\Theta}}(\hat{A_{\mu}})$, where $\phi_{ijk}=\chi_{ijk}=\psi_{i}\psi_{j}\psi_{k}$. See table \ref{TChA}.

\begin{table}[h]
\begin{center}
\resizebox{1.0\textwidth}{!}{
\begin{tabular}{cccccccccccccccccccc}
\hline
\hline
IIR's (Ch) & \vline \vline & $\hat{I}$ & $\hat{C}$ & $\hat{P}$ & $\hat{T}$ & $\hat{P}*\hat{T}$ & $\hat{C}*\hat{P}$ & $\hat{C}*\hat{T}$  & $\hat{\Theta}$\\
$G_{\hat{\Theta}}(\hat{A_{\mu}})$ & \vline \vline & $(e_{1},e_{2},e_{3})$ & $(a_{1},e_{2},e_{3})$ & $(e_{1},a_{2},e_{3})$ & $(e_{1},e_{2},a_{3})$ & $(e_{1},a_{2},a_{3})$ & $(a_{1},a_{2},e_{3})$ & $(a_{1},e_{2},a_{3})$ & $(a_{1},a_{2},a_{3})$ \\
\hline
\hline
$\phi_{111}=\Phi_{1}$ & \vline \vline & $1$ & $1$ & $1$ & $1$ & $1$ & $1$ & $1$ & $1$ \\
$\phi_{211}=\Phi_{2}$ & \vline \vline & $1$ & $-1$ & $1$ & $1$ & $1$ & $-1$ & $-1$ & $-1$ \\
$\phi_{121}=\Phi_{3}$ & \vline \vline & $1$ & $1$ & $-1$ & $1$ & $-1$ & $-1$ & $1$ & $-1$ \\
$\phi_{112}=\Phi_{4}$ & \vline \vline & $1$ & $1$ & $1$ & $-1$ & $-1$ & $1$ & $-1$ & $-1$ \\
$\phi_{221}=\Phi_{5}$ & \vline \vline & $1$ & $-1$ & $-1$ & $1$ & $-1$ & $1$ & $-1$ & $1$ \\
$\phi_{212}=\Phi_{6}$ & \vline \vline & $1$ & $-1$ & $1$ & $-1$ & $-1$ & $-1$ & $1$ & $1$ \\
$\phi_{122}=\Phi_{7}$ & \vline \vline & $1$ & $1$ & $-1$ & $-1$ & $1$ & $-1$ & $-1$ & $1$ \\
$\phi_{222}=\Phi_{8}$ & \vline \vline & $1$ & $-1$ & $-1$ & $-1$ & $1$ & $1$ & $1$ & $-1$ \\
\hline
\hline
\end{tabular}}
\end{center}
\caption[]{\small{IIR's (characters) table of $G_{\hat{\Theta}}(\hat{A_{\mu}})$.}}
\label{TChA}
\end{table}
\vspace{0.5cm}

{\centering \section{IIR's of $\mathbf{G_{\bm \Theta}^{(2)}(\bm \psi)}$}}

\hspace{18pt} In \cite{CPT group} it was shown that at the level of the Dirac equation, for the 4-spinor $\psi$, exists two $CPT$ groups, $G_{\Theta}^{(1)}(\psi)$ and $G_{\Theta}^{(2)}(\psi)$, whose elements are constructed with products of Dirac $\gamma$-matrices. Only $G_{\Theta}^{(2)}(\psi)$, which turns out isomorphic to the semi-direct product $D_{4}\rtimes \mathbb{Z}_{2}$, where $D_{4}\equiv DH_{8}$ is the dihedral group of eight elements (the group of symmetries of the square), is compatible with $G_{\hat{\Theta}}(\hat{\psi})$.

It is then of interest to construct the IIR's of $G_{\Theta}^{(2)}(\psi)$, both for completness and also for comparison with the IIR's of the operator group. 

The action of $\mathbb{Z}_{2}\cong\left\{ 1,-1 \right\}$ on $D_{4}$ as a subgroup of $S_{4}$ (the symmetric group of 4 elements) is given by

\begin{eqnarray}
\lambda: \mathbb{Z}_{2}\rightarrow Aut(D_{4}),\nonumber\\
1\mapsto \lambda(1)&=&Id_{D_{4}},\nonumber\\
\lambda(-1)(I)&=&I,\nonumber\\
\lambda(-1)(1234)&=&(1234),\nonumber\\
\lambda(-1)(24)&=&(13),\nonumber\\
\lambda(-1)(13)&=&(24),\nonumber\\
\lambda(-1)((12)(34))&=&(14)(23),\nonumber\\
\lambda(-1)((14)(23))&=&(12)(34),\nonumber\\
\lambda(-1)((13)(24))&=&(13)(24),\nonumber\\
\lambda(-1)(1432)&=&(1432).
\end{eqnarray}

Here
\begin{equation}
D_{4}=\left\{ I; (1234), (1432); (13)(24); (12)(34), (14)(23); (24),(13) \right\},
\label{D4}
\end{equation}
where between semi-colons we have enclosed the five conjugation classes.

Then, the composition law in $D_{4}\rtimes \mathbb{Z}_{2}$ is

\begin{equation}
(g',h')(g,h)=(g' \lambda(h')(g),h'h):
\end{equation}
\begin{eqnarray}
(g',1)(g,h)&=&(g'g,h)\\
(g',-1)(g,h)&=&(g' \lambda(-1)(g),-h),	
\end{eqnarray}
where, as a set,

\begin{eqnarray}
D_{4}\times \mathbb{Z}_{2}=\left\{ (I,1), (I,-1),(-I,1), (-I,-1), (P,1),(P,-1), (-P,1), \right. \nonumber\\
\left. (-P,-1),(CT,1), (CT,-1), (-CT,1), (-CT,-1),(\Theta,1), \right. \nonumber\\
\left. (\Theta,-1), (-\Theta,1), (-\Theta,-1) \right\},	
\end{eqnarray}

with inverses
\begin{equation}
(g,h)^{-1}=(\lambda(h)(g^{-1}),h).
\end{equation}

From eq. (55) in \cite{CPT group}, there is the isomorphism
\begin{equation}
D_{4}\rightarrow \left\{ I, -I, P, -P, CT, -CT, \Theta, -\Theta \right\} 
\label{D4CPT}
\end{equation}
given by
\begin{eqnarray}
I\mapsto I,\nonumber\\
(1234)\mapsto P,\nonumber\\
(1432)\mapsto -P,\nonumber\\
(13)(24)\mapsto -I,\nonumber\\
(12)(34)\mapsto \Theta,\nonumber\\
(14)(23)\mapsto -\Theta,\nonumber\\
(24)\mapsto -CT,\nonumber\\
(13)\mapsto CT
\label{isoD4CPT}
\end{eqnarray}
while from eq. (60) in \cite{CPT group} and eq. (\ref{D4}), the isomorphism between $G_{\Theta}^{(2)}$ and $D_{4}\rtimes \mathbb{Z}_{2}$,
\begin{equation}
G_{\Theta}^{(2)}\rightarrow D_{4}\rtimes \mathbb{Z}_{2},
\end{equation}
is given by:

\begin{eqnarray}
I\mapsto (I,1),\qquad -I\mapsto (-I,1),\nonumber\\
C\mapsto (-\Theta,-1),\qquad -C\mapsto (\Theta,-1),\nonumber\\
P\mapsto (P,1),\qquad -P\mapsto (-P,1),\nonumber\\
T\mapsto (P,-1),\qquad -T\mapsto (-P,-1),\nonumber\\
CP\mapsto (CT,-1),\qquad -CP\mapsto (-CT,-1),\nonumber\\
CT\mapsto (CT,1),\qquad -CT\mapsto (-CT,1),\nonumber\\
PT\mapsto (-I,-1),\qquad -PT\mapsto (I,-1),\nonumber\\
\Theta \mapsto (\Theta,1),\qquad -\Theta \mapsto (-\Theta,1).
\end{eqnarray}

It is easy to verify that $G_{\Theta}^{(2)}$ has ten conjugation classes, namely
\begin{eqnarray}
\text{$[$}I]&=&=\lbrace I \rbrace,\nonumber\\
\text{$[$}-I]&=&=\lbrace -I \rbrace,\nonumber\\
\text{$[$}C]&=&=\lbrace C,-C \rbrace,\nonumber\\
\text{$[$}T]&=&=\lbrace T \rbrace,\nonumber\\
\text{$[$}-T]&=&=\lbrace -T \rbrace,\nonumber\\
\text{$[$}P]&=&=\lbrace P,-P \rbrace,\nonumber\\
\text{$[$}CP]&=&=\lbrace CP,-CP \rbrace,\nonumber\\
\text{$[$}CT]&=&=\lbrace CT,-CT \rbrace,\nonumber\\
\text{$[$}PT]&=&=\lbrace PT,-PT \rbrace,\nonumber\\
\text{$[$}\Theta]&=&=\lbrace \Theta,-\Theta \rbrace.
\end{eqnarray}

Then, being a finite group, $G_{\Theta}^{(2)}$ has as many IIR's as conjugation classes. Since the sum of the squares of the dimensions of these representations must be 10, a simple calculation leads to the existence of eight 1-dimensional irrep's $\varphi_{k}$, $k=1,...8$, and two 2-dimensional irrep's, $\varphi_{9}$ and $\varphi_{10}$. 

$\varphi_{1}$ is the trivial representation $g \mapsto 1$, for all $g\,\in G_{\Theta}^{(2)}$. The three 1-dimensional irrep's $\varphi_{2}$, $\varphi_{3}$ and $\varphi_{4}$ are obtained taking into account:

\begin{enumerate}
\item $D_{4}$, $C_{4}\times \mathbb{Z}_{2}$ and $Q$ are invariant subgroups of $D_{4}\rtimes \mathbb{Z}_{2}$.
\item The well known theorem by which if $H$ is an invariant subgroup of $G$ and $\varphi: G/H\rightarrow K<GL(V)$ is a representation of $G/H$ over the vector space $V$, then $\varphi\circ p$ is a degenerate representation of $G$, where $p: G\rightarrow G/H$ is the canonical projection $p(g)=<g>=gH$ \cite{Tung}. Thus, we have the commutative diagram:
\begin{equation}
\begin{tabular}{ccccc}
 &  & K &  &  \\ 
 & $\varphi \circ 	p \nearrow$ & $\:\:  $ & $\:\nwarrow \varphi$ \\ 
 & G &$\underset{p}{\longrightarrow}$  & $G/H $   \\ 
\end{tabular} .
\label{triangulito}
\end{equation}
\end{enumerate}

$D_{4}$ is given by the r.h.s. of  eq. (\ref{D4CPT}) and $C_{4}=<\left\{ P \right\}>$, so
\begin{equation}
C_{4}\times \mathbb{Z}_{2}=\left\{ (I,1), (I,-1),(-I,1), (-I,-1), (P,1), (P,-1), (-P,1), (-P,-1)\right\},
\end{equation}
and 
\begin{equation}
Q\cong <\left\{ C, P \right\} >=\left\{ I, C, P, CP,-I -C, -P, -CP \right\}.
\end{equation}

In diagram (\ref{triangulito}), we choose $G=D_{4}\rtimes \mathbb{Z}_{2}$ and $H$ equal to $D_{4}$, $C_{4}\times \mathbb{Z}_{2}$ and $Q$, respectively. Then

\begin{equation}
\frac{D_{4}\rtimes \mathbb{Z}_{2}}{H}	\cong \mathbb{Z}_{2}=\left\{ e, a \right\}
\end{equation}
with
\begin{eqnarray}
e&=&D_{4}, \qquad a=\left\{C, T, CP, PT, -C, -T, -CP, -PT \right\},\nonumber\\
e&=&C_{4}\times \mathbb{Z}_{2}, \qquad a=\left\{C, CP, CT, \Theta, -C, -CP, -CT, -\Theta \right\},\nonumber\\
e&=&Q, \qquad a=\left\{T, CT, PT, \Theta, -T, -CT, -PT, -\Theta \right\},
\end{eqnarray}
respectively. The elements in the $e$'s are represented by $1$, while the elements in the $a$'s are represented by $-1$ (see table \ref{TIDE}).

As is well known, $D_{4}$ has five IIR's (one 2-dimensional and four 1-dimensional). The 2-dimensional irrep is given by

\begin{eqnarray}
I\mapsto \begin{pmatrix} 1 & 0 \cr 0 & 1 \end{pmatrix}=1_{2\times 2}\equiv I\nonumber\\
(13)(24)\mapsto \begin{pmatrix} -1 & 0 \cr 0 & -1 \end{pmatrix}=-1_{2\times 2}\equiv -I\nonumber\\
(1234)\mapsto \begin{pmatrix} 0 & -1 \cr 1 & 0 \end{pmatrix}=-i\sigma_{2}\equiv P\nonumber\\
(1432)\mapsto \begin{pmatrix} 0 & 1 \cr -1 & 0 \end{pmatrix}=i\sigma_{2}\equiv -P\nonumber\\
(12)(34)\mapsto \begin{pmatrix} -1 & 0 \cr 0 & 1 \end{pmatrix}=-\sigma_{3}\equiv \Theta\nonumber\\
(14)(23)\mapsto \begin{pmatrix} 1 & 0 \cr 0 & -1 \end{pmatrix}=\sigma_{3}\equiv -\Theta\nonumber\\
(24)\mapsto \begin{pmatrix} 0 & 1 \cr 1 & 0 \end{pmatrix}=\sigma_{1}\equiv -CT \nonumber\\
(13)\mapsto \begin{pmatrix} 0 & -1 \cr -1 & 0 \end{pmatrix}=-\sigma_{1}\equiv CT,
\end{eqnarray}
where in the last identification we used the relations (\ref{isoD4CPT}).

With the choice $C=i\sigma_{1}$, we obtain,
\begin{equation}
T=\begin{pmatrix} i & 0 \cr 0 & i \end{pmatrix}=iI,\quad CP=\begin{pmatrix} i & 0 \cr 0 & -i \end{pmatrix}=i\sigma_{3},\quad PT=\begin{pmatrix} 0 & -i \cr i & 0 \end{pmatrix}=\sigma_{2},
\end{equation}
which completes the 2-dimensional irrep $\varphi_{9}$ of $G_{\Theta}^{(2)}$; while with the choice $C=-i\sigma_{1}$, we obtain,
\begin{equation}
T=\begin{pmatrix} -i & 0 \cr 0 & -i \end{pmatrix}=-iI,\quad CP=\begin{pmatrix} -i & 0 \cr 0 & i \end{pmatrix}=-i\sigma_{3},\quad PT=\begin{pmatrix} 0 & i \cr -i & 0 \end{pmatrix}=-\sigma_{2},
\end{equation}
which completes the 2-dimensional irrep $\varphi_{10}$ of $G_{\Theta}^{(2)}$. It can be easily verified that $\varphi_{9}$ and $\varphi_{10}$ are inequivalent, i.e., it does not exist a matrix $S$ such that $SMS^{\dagger}=M'$ (or $SMS^{-1}=M'$) for $M\,\in\, \varphi_{9}$ and $M'\,\in\, \varphi_{10}$.

The remaining four 1-dimensional IIR's, $\varphi_{5}$, $\varphi_{6}$, $\varphi_{7}$ and $\varphi_{8}$, are obtained from the orthogonality between the columns of the characters table for conjugation classes (completeness relation). Then, the orthogonality between the rows of the complete characters table is verified.

The IIR's and characters for $G_{\Theta}^{(2)}(\psi)\cong D_{4}\rtimes \mathbb{Z}_{2}$ are summarized in tables \ref{TIDE} and \ref{TChDE}. Comparing with the tables for $G_{\hat{\Theta}}(\hat{\psi})$, we see that the 1-dimensional irrep's coincide:

\begin{eqnarray}
\phi_{1}=\varphi_{1},\; \phi_{2}=\varphi_{3},\;\phi_{3}=\varphi_{4},\; \phi_{4}=\varphi_{2},\nonumber\\
\phi_{5}=\varphi_{8},\; \phi_{6}=\varphi_{6},\;\phi_{7}=\varphi_{7},\; \phi_{8}=\varphi_{5}.
\end{eqnarray}
However, $\phi_{9}$ is not equivalent to either $\varphi_{9}$ or $\varphi_{10}$ and the same holds for $\phi_{10}$.

\begin{table}[h]
\begin{center}
\begin{sideways}
\resizebox{1.5\textwidth}{!}{
\begin{tabular}{cccccccccccccccccccc}
\hline
\hline
IIR's $G_{\Theta}^{(2)}(\psi)$ & \vline \vline & $I$ & $C$ & $-I$ & $-C$ & $P$ & $CP$ & $-P$ & $-CP$ & $T$ & $CT$ & $-T$ & $-CT$ & $PT$ & $\Theta$ & $-PT$ & $-\Theta$\\
\hline
\hline
$\varphi_{1}$ & \vline \vline & $1$ & $1$ & $1$ & $1$ & $1$ & $1$ & $1$ & $1$ & $1$ & $1$ & $1$ & $1$ & $1$ & $1$ & $1$ & $1$\\
$\varphi_{2}$ & \vline \vline & $1$ & $-1$ & $1$ & $-1$ & $1$ & $-1$ & $1$ & $-1$ & $-1$ & $1$ & $-1$ & $1$ & $-1$ & $1$ & $-1$ & $1$ \\
$\varphi_{3}$ & \vline \vline & $1$ & $-1$ & $1$ & $-1$ & $1$ & $-1$ & $1$ & $-1$ & $1$ & $-1$ & $1$ & $-1$ & $1$ & $-1$ & $1$ & $-1$\\
$\varphi_{4}$ & \vline \vline & $1$ & $1$ & $1$ & $1$ & $1$ & $1$ & $1$ & $1$ & $-1$ & $-1$ & $-1$ & $-1$ & $-1$ & $-1$ & $-1$ & $-1$\\
$\varphi_{5}$ & \vline \vline & $1$ & $-1$ & $1$ & $-1$ & $-1$ & $1$ & $-1$ & $1$ & $-1$ & $1$ & $-1$ & $1$ & $1$ & $-1$ & $1$ & $-1$\\
$\varphi_{6}$ & \vline \vline & $1$ & $-1$ & $1$ & $-1$ & $-1$ & $1$ & $-1$ & $1$ & $1$ & $-1$ & $1$ & $-1$ & $-1$ & $1$ & $-1$ & $1$\\
$\varphi_{7}$ & \vline \vline & $1$ & $1$ & $1$ & $1$ & $-1$ & $-1$ & $-1$ & $-1$ & $-1$ & $-1$ & $-1$ & $-1$ & $1$ & $1$ & $1$ & $1$\\
$\varphi_{8}$ & \vline \vline & $1$ & $-$ & $1$ & $-$ & $-1$ & $-1$ & $-1$ & $-1$ & $1$ & $1$ & $1$ & $1$ & $-1$ & $-1$ & $-1$ & $-1$\\
$\varphi_{9}$ & \vline \vline & $\begin{pmatrix} 1 & 0 \cr 0 & 1 \end{pmatrix}$ & $\begin{pmatrix} 0 & i \cr i & 0 \end{pmatrix}$ & $\begin{pmatrix} -1 & 0 \cr 0 & -1 \end{pmatrix}$ & $\begin{pmatrix} 0 & -i \cr -i & 0 \end{pmatrix}$ & $\begin{pmatrix} 0 & -1 \cr 1 & 0 \end{pmatrix}$ & $\begin{pmatrix} i & 0 \cr 0 & -i \end{pmatrix}$ & $\begin{pmatrix} 0 & 1 \cr -1 & 0 \end{pmatrix}$ & $\begin{pmatrix} -i & 0 \cr 0 & i \end{pmatrix}$ & $\begin{pmatrix} i & 0 \cr 0 & i \end{pmatrix}$ & $\begin{pmatrix} 0 & -1 \cr -1 & 0 \end{pmatrix}$ & $\begin{pmatrix} -i & 0 \cr 0 & -i \end{pmatrix}$ & $\begin{pmatrix} 0 & 1 \cr 1 & 0 \end{pmatrix}$ & $\begin{pmatrix} 0 & -i \cr i & 0 \end{pmatrix}$ & $\begin{pmatrix} -1 & 0 \cr 0 & 1 \end{pmatrix}$ & $\begin{pmatrix} 0 & i \cr -i & 0 \end{pmatrix}$ & $\begin{pmatrix} 1 & 0 \cr 0 & -1 \end{pmatrix}$\\
$\varphi_{10}$ & \vline \vline & $\begin{pmatrix} 1 & 0 \cr 0 & 1 \end{pmatrix}$ & $\begin{pmatrix} 0 & -i \cr -i & 0 \end{pmatrix}$ & $\begin{pmatrix} -1 & 0 \cr 0 & -1 \end{pmatrix}$ & $\begin{pmatrix} 0 & i \cr i & 0 \end{pmatrix}$ & $\begin{pmatrix} 0 & -1 \cr 1 & 0 \end{pmatrix}$ & $\begin{pmatrix} -i & 0 \cr 0 & i \end{pmatrix}$ & $\begin{pmatrix} 0 & 1 \cr -1 & 0 \end{pmatrix}$ & $\begin{pmatrix} i & 0 \cr 0 & -i \end{pmatrix}$ & $\begin{pmatrix} -i & 0 \cr 0 & -i \end{pmatrix}$ & $\begin{pmatrix} 0 & -1 \cr -1 & 0 \end{pmatrix}$ & $\begin{pmatrix} i & 0 \cr 0 & i \end{pmatrix}$ & $\begin{pmatrix} 0 & -1 \cr -1 & 0 \end{pmatrix}$ & $\begin{pmatrix} 0 & i \cr -i & 0 \end{pmatrix}$ & $\begin{pmatrix} -1 & 0 \cr 0 & 1 \end{pmatrix}$ & $\begin{pmatrix} 0 & -i \cr i & 0 \end{pmatrix}$ & $\begin{pmatrix} 1 & 0 \cr 0 & -1 \end{pmatrix}$\\
\hline
\hline
\end{tabular}}
\end{sideways}
\end{center}
\caption[]{\small{IIR's table of $G_{\Theta}^{(2)}(\psi)$.}}
\label{TIDE}
\end{table}

\begin{table}[h]
\begin{center}
\resizebox{1.0\textwidth}{!}{
\begin{tabular}{cccccccccccccccccccc}
\hline
\hline
Ch $G_{\Theta}^{(2)}(\psi)$ & \vline \vline & $[I]$ & $[-I]$ & $2[C]$ & $[T]$ & $[-T]$ & $2[P]$ & $2[CP]$ & $2[CT]$ & $2[PT]$ &  $2[\Theta]$\\
\hline
\hline
$\chi_{1}$ & \vline \vline & $1$ & $1$ & $1$ & $1$ & $1$ & $1$ & $1$ & $1$ & $1$ & $1$\\
$\chi_{2}$ & \vline \vline & $1$ & $1$ & $-1$ & $-1$ & $-1$ & $1$ & $-1$ & $1$ & $-1$ & $1$\\
$\chi_{3}$ & \vline \vline & $1$ & $1$ & $-1$ & $1$ & $1$ & $1$ & $-1$ & $-1$ & $1$ & $-1$\\
$\chi_{4}$ & \vline \vline & $1$ & $1$ & $1$ & $-1$ & $-1$ & $1$ & $1$ & $-1$ & $-1$ & $-1$\\
$\chi_{5}$ & \vline \vline & $1$ & $1$ & $-1$ & $-1$ & $-1$ & $-1$ & $1$ & $1$ & $1$ & $-1$\\
$\chi_{6}$ & \vline \vline & $1$ & $1$ & $-1$ & $1$ & $1$ & $-1$ & $1$ & $-1$ & $-1$ & $1$\\
$\chi_{7}$ & \vline \vline & $1$ & $1$ & $1$ & $-1$ & $-1$ & $-1$ & $-1$ & $-1$ & $1$ & $1$\\
$\chi_{8}$ & \vline \vline & $1$ & $1$ & $1$ & $1$ & $1$ & $-1$ & $-1$ & $1$ & $-1$ & $-1$\\
$\chi_{9}$ & \vline \vline & $2$ & $-2$ & $0$ & $2i$ & $-2i$ & $0$ & $0$ & $0$ & $0$ & $0$\\
$\chi_{10}$ & \vline \vline & $2$ & $-2$ & $0$ & $-2i$ & $2i$ & $0$ & $0$ & $0$ & $0$ & $0$\\
\hline
\hline
\end{tabular}}
\end{center}
\caption[]{\small{Characters table of $G_{\Theta}^{(2)}(\psi)$.}}
\label{TChDE}
\end{table}

{\centering \section{Final comments}}

\hspace{18pt} It is of interest to ask the question if, from the point of view of group theory, the behaviour of the photon field $\hat{A_{\mu}}$ under the $C, P, T$ transformation is independent of the behaviour of the electron-positron field $\hat{\psi}$. The answer to this question is afirmative. In fact, in $G_{\hat{\Theta}}(\hat{\psi})$ there is only one $\mathbb{Z}_{2}$ subgroup, namely that generated by $\hat{C}: \:\mathbb{Z}_{2}\cong\lbrace \hat{I}, \hat{C} \rbrace$. Then, $G_{\hat{\Theta}}(\hat{A_{\mu}})\cong\mathbb{Z}_{2}^{3}$ is not a subgroup of $G_{\hat{\Theta}}(\hat{\psi})$:

\begin{equation}
G_{\hat{\Theta}}(\hat{A_{\mu}}) \not<  G_{\hat{\Theta}}(\hat{\psi}).
\end{equation}

The same happens in relation to $G_{\Theta}^{(2)}(\psi)$. In this group there are three $\mathbb{Z}_{2}$-subgroups:
\begin{equation}
\mathbb{Z}_{2}^{(1)}=\lbrace I, CT \rbrace, \; \mathbb{Z}_{2}^{(2)}=\lbrace I, PT \rbrace, \; \mathbb{Z}_{2}^{(3)}=\lbrace I, \Theta \rbrace.
\end{equation}

However, the table of $\mathbb{Z}_{2}^{3}$ has eight $1$'s, while the table of $G_{\Theta}^{(2)}(\psi)$ has only seven $1$'s. So

\begin{equation}
G_{\hat{\Theta}}(\hat{A_{\mu}})\not<G_{\Theta}^{(2)}.
\end{equation}

Finally, in relation with the posible relevance of the $CPT$ group structures in other areas of physics, besides field theory, we call the attention on the recently found relation between the groups $G_{\hat{\Theta}}(\hat{\psi})\cong Q\times\mathbb{Z}_{2}$, $G_{\Theta}^{(2)}(\psi)\cong D_{4}\rtimes \mathbb{Z}_{2}$ and $G_{\hat{\Theta}}(\hat{A_{\mu}})\cong \mathbb{Z}_{2}^{3}$, and fundamental theorems in the context of quantum paradoxes, like the Kochen-Specker theorem \cite{M}.

{\centering \section*{Acknowledgments}}

\hspace{18pt} This work was partially support by the project PAPIIT IN 118609-2, DGAPA-UNAM, M\'exico. B. Carballo P\'erez also acknowledge financial support from CONACyT, M\'exico. The authors thank Luis Perissinotti for useful comments.

\newpage
\lhead{} 
\rhead{}

\end{document}